\documentclass[doublecol]{epl2}

\title{Torque anomalies at magnetization plateaux in quantum magnets
  with Dzyaloshinskii-Moriya interactions} 
\author{S.R. Manmana \and F. Mila}
\institute{Institute of Theoretical Physics, \'Ecole
  Polytechnique F\'ed\'erale de Lausanne, CH-1015 Lausanne,
  Switzerland}

\date{\today}

\abstract{
We investigate the effect of Dzyaloshinskii-Moriya (DM) interactions
on torque measurements of quantum magnets with magnetization plateaux
in the context of a frustrated spin-1/2 ladder.  
Using extensive DMRG simulations, we show that the DM contribution to
the torque is peaked at the critical fields, 
and that the total torque is non-monotonous if the DM interaction is
large enough compared to the g-tensor anisotropy.  
More remarkably, if the DM vectors point in a principal direction of
the g-tensor, torque measurements close to this direction will show
well defined peaks even for small DM interaction, leading to a very
sensitive way to detect the critical fields. 
We propose to test this effect in the two-dimensional plateau system
SrCu$_2$(BO$_3$)$_2$. }

\begin{document}

\maketitle

The investigation of quantum magnets in high magnetic fields is a very 
active field of research thanks to a number of recent and remarkable
discoveries ranging from Bose-Einstein condensation\cite{ruegg} to magnetization
plateaux\cite{kageyama,kodama}, and to the on-going search for the analog of supersolid
phases\cite{ng,batista,laflorencie,schmidt}. 
A central piece of information is provided by the magnetization as a
function of the external field. However, technical requirements have lead
many groups to measure the torque rather than the magnetization. 
In SU(2) invariant magnets, it does not matter since the torque is
rigorously proportional to the magnetization along the field. 
In the presence of anisotropic interactions, such as Dzyaloshinskii-Moriya (DM) interactions\cite{DMoriginal}, 
this is no longer the case and an additional response is obtained which needs to be considered carefully.
This DM contribution to the torque has already been
investigated in the context of molecular magnets\cite{Cinti,Lante} and of the spin
ladder compound Cu(Hp)Cl\cite{clemancey}.  
In the latter case, the DM interactions were assumed not to compete
with the exchange, 
%in which case 
so that their additional effect on the torque is a relatively smooth contribution in the intermediate phase. \\
Recent torque measurements on SrCu$_2$(BO$_3$)$_2$, a quasi-two-dimensional
quantum magnet with several magnetization plateaux\cite{kageyama,kodama}, call for further
investigation of the issue. 
In particular, the field dependence of the torque reported in
refs. \cite{torque_Sebastian,torque_Levy} is quite different, 
especially regarding the behavior inside the plateaux. 
In Sebastian {\it et al.}'s data, the torque is never flat but increases quite
significantly inside the plateaux, while in Levy {\it et al.}'s data, it actually 
{\it decreases} inside the 1/8 plateau. 
Since DM interactions have been unambiguously identified in
SrCu$_2$(BO$_3$)$_2$\cite{cepas,zorko,kodama2}, the torque 
is not expected to be simply
proportional to the magnetization, which leaves the door open to
%different 
anomalous 
behavior depending on the details of the experiments, in
particular the orientation of the field. 
However, the field dependence of the torque around magnetization
plateaux has never been investigated so far. 

\begin{figure}[t]
\includegraphics[width=0.49\textwidth]{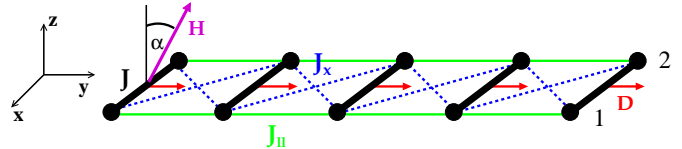}
\caption{(Colour on-line) Ladder system under consideration. 
Red arrows indicate DM vectors, the magnetic field lies in the
$(y,z)$ plane, but is tilted by an angle $\alpha$ with respect to the
$z$-axis.} 
\label{Fig:01}
\end{figure}

In this Letter, we investigate this issue in the context of the
frustrated ladder depicted in fig. 1 and defined by the Hamiltonian
($\hbar \equiv 1$)
\begin{eqnarray}
\mathcal{H} &=& J \sum\limits_j \vec{S}_{j,1} \cdot \vec{S}_{j,2} \nonumber
            + \sum\limits_j  \vec{D}_j \cdot \left(
            \vec{S}_{j,1} \times \vec{S}_{j,2} \right)
            \\ \nonumber 
        &&  + J_{\parallel} \sum\limits_j 
        \left( \vec{S}_{j,1} \cdot \vec{S}_{j+1,1} 
               + \vec{S}_{j,2} \cdot \vec{S}_{j+1,2} \right) \qquad\\ \nonumber 
        &&  + J_{\times} \sum\limits_j 
        \left( \vec{S}_{j,1} \cdot \vec{S}_{j+1,2} 
               + \vec{S}_{j,2} \cdot \vec{S}_{j+1,1} \right) \qquad \\ 
        &&  - N \mu_B \vec{H} \mathbf{g} \vec{S}, \\
\vec{S} &=& \frac{1}{N}\sum\limits_j \left( \vec{S}_{j,1} +
\vec{S}_{j,2} \right) 
\label{Eq:Hamiltonian}
\end{eqnarray}
This frustrated ladder has a magnetization plateau at 1/2 if 
$J_{\parallel}$ and $J_{\times}$ are not too different (their ratio should
be between 1/3 and 3 in the limit $J_{\parallel},J_{\times}\ll J$)\cite{Mila}, and it
is amenable to the density matrix renormalization group
method (DMRG). It is thus an ideal minimal model to study subtle
effects related to plateaux in frustrated quantum magnets.
We set $J \equiv 1$ throughout the paper and we will concentrate
on the case $J_{\parallel}=J_{\times}$ for which the plateau is largest.
We assume the g-tensor $\mathbf{g}$ to be diagonal but to possess an
asymmetry $\delta g$ between the $x$-$y$ and the $z$ component. 
Due to the DM interaction, the SU(2) symmetry of the original
Heisenberg model is  not present and the DMRG calculations can be very
demanding, restricting us to treat only rather small systems with up
to 49 rungs.     

\begin{figure}[t]
\includegraphics[width=0.49\textwidth]{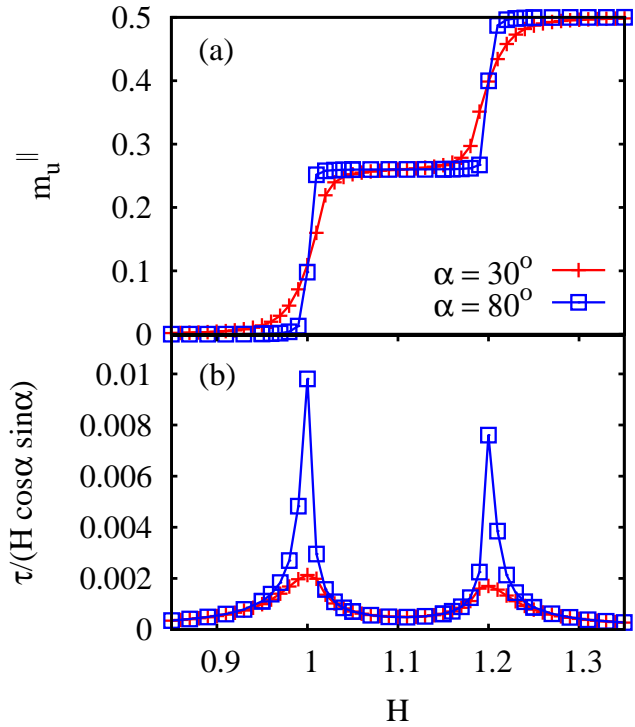}
\caption{(Colour on-line) DMRG results for the uniform magnetization (a) parallel and
  (b) perpendicular to $\vec{H}$ for $D = 0.03$ and $\delta g = 0$
  for $\alpha = 30^{\circ}$ or $80^{\circ}$, respectively, and $J_{\parallel} = J_{\times} = 0.1$ for systems with 25 rungs. 
  In (b), the magnetization is scaled by $\sin \alpha \, \cos \alpha $,
  the angular dependence of the torque of a single dimer with DM
  anisotropy away from the critical field.} 
\label{Fig:02}
\end{figure}

To supplement the finite size DMRG data, we have also performed
a mean-field calculation which generalizes the ansatz used in ref. \cite{PRL_Karlo} in order
to deal with the tilted field. It is based on a product of rung wave functions 
\begin{eqnarray}
| \psi \rangle &=& \prod_j | \phi \rangle_j \nonumber \\
| \phi \rangle_j &=& a |S \rangle_j + b |T_{-1} \rangle_j + c | T_{0}
| \rangle_j + d | T_1 \rangle_j
\label{eq:meanfield}
\end{eqnarray}
with the singlet and triplet eigenstates of the dimer on rung $j$
given by $|S \rangle_j = 1/\sqrt{2} (| \downarrow \uparrow \rangle_j - |
\uparrow \downarrow \rangle_j), \, |T_{-1}~\rangle_j = | \downarrow
\downarrow \rangle_j, \, |T_0 \rangle_j = 1/\sqrt{2} (|\downarrow
\uparrow \rangle_j + | \uparrow \downarrow \rangle_j),$ and $|T_1
\rangle_j = | \uparrow \uparrow \rangle_j$.
For the case $J_{\parallel} = J_{\times}$ and $D=0$, this ansatz
provides an exact solution \cite{Mila}, and it is expected to provide a good
approximation for the physically relevant case $D \ll J$.
Comparison with the finite system DMRG data shows good agreement, 
giving us confidence that the results presented are not affected
by strong finite size effects. 

In the following we focus on the magnetic response of the system to an
external magnetic field with arbitrary orientation. 
First we discuss our results in the presence of $D$ only, 
and afterwards go to the more general case with an additional finite 
$\delta g$. 
The symmetry analysis for a single dimer with DM interaction and
$\delta g = 0$ ascertains that the results are independent of the
absolute orientation of $\vec{H}$ and $\vec{D}$ as long as the angle
between them is the same\cite{PRB_Shin} for which reason we confine
ourselves to treat only the case depicted in fig. \ref{Fig:01}. 
If not mentioned otherwise, we present results for $D = 0.03$ and
$J_{\parallel} = J_{\times} = 0.1$, i.e., we restrict ourselves to the
case of a strongly frustrated ladder, which is relevant when having in
mind the frustrated plateau system SrCu$_2$(BO$_3$)$_2$.
\begin{figure}[t]
\includegraphics[width=0.49\textwidth]{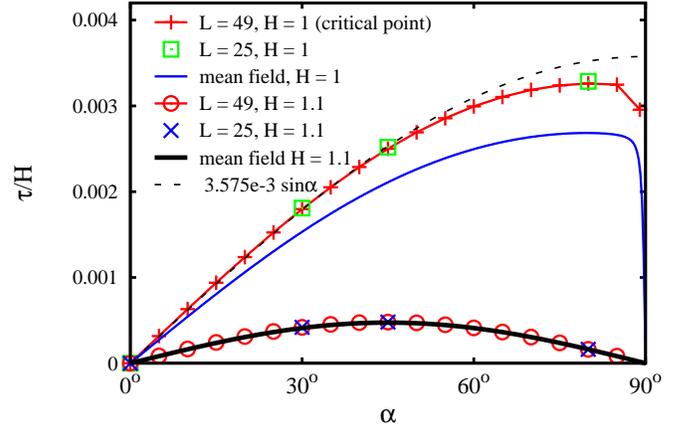}
\caption{(Colour on-line) Angular dependence of $\tau/H$ at the critical field $H = 1$ and inside the
  plateau ($H = 1.1$) as obtained from DMRG calculations with $L =25$ and
  $L=49$ rungs and from the mean-field calculation using the ansatz
  eq. (\ref{eq:meanfield}). 
%  The $\sin \alpha $ and $\sin \alpha  \, \cos \alpha$ functions
 % indicate the angular dependence of an isolated dimer with DM
 % anisotropy.
% For $H=1.1$, the results are in very good agreement with the fit $ 9.2 \cdot 10^{-4} \sin \alpha \, \cos \alpha$, while for $H=1$ for angles larger than $45^{\circ}$ the DMRG results deviate from the fit $3.575 \cdot 10^{-3} \sin \alpha$.
 }  
\label{Fig:03}
\end{figure}
We concentrate on the uniform magnetization, which is defined as  
\begin{equation}
\vec{m}_u  := - \frac{1}{N} \left\langle \frac{\partial
  \mathcal{H}}{\partial \vec{H}} \right\rangle = \mu_B \mathbf{g} \,
\left\langle \vec{S} \right\rangle. 
\end{equation}
In general, for finite $\delta g$ a magnetization perpendicular to
$\vec{H}$ is induced, causing a torque 
\begin{equation}
\vec{\tau} = \vec{m}_u \times \vec{H},
\end{equation}
which gives access to the uniform magnetization perpendicular to
$\vec{H}$ obtained as $m_u^{\perp} = \tau/H$.  
If $D = 0$, in order to minimize the Zeeman-term in the Hamiltonian,
the spins align in the direction of the vector $\vec{H} \mathbf{g}$. 
For a better comparison of our results obtained at various angles and for
different values of $\delta g$, let us introduce the effective field 
\begin{equation}
\vec{H}^{\rm eff} := \mu_B \vec{H} \mathbf{g}.
\end{equation}
In the following, when $\delta g \neq 0$, we discuss the
dependence $\vec{m}_u(H^{\rm eff})$.
If $D = 0$ and $\delta g \neq 0$, the components of the uniform
magnetization parallel and perpendicular to $\vec{H}$ are obtained as
\begin{eqnarray}
\vec{m}_u^{\parallel} &=&  \left|\left\langle \vec{S} \right\rangle
\right| \mu_B  \sqrt{g^2 + \cos^2\alpha \, \delta g (2g + \delta g)}
%\sin^2\alpha g^2 + \cos^2\alpha (g+\delta g)^2} 
\left(
\begin{array}{c}
0\\
\sin \alpha\\
\cos\alpha
\end{array}
\right)\quad\ \\
%\qquad \frac{\vec{H}}{H}\\
\vec{m}_u^{\perp} &=&  \left|\left\langle \vec{S} \right\rangle \right| 
\frac{\mu_B \sin \alpha \cos \alpha \, \delta g (2g +  \delta g) }
{\sqrt{g^2 + \cos^2\alpha \, \delta g (2g + \delta g)}}
%{\sqrt{\sin^2 \alpha g^2 + \cos^2 \alpha (g+\delta g)^2}}
\left(
\begin{array}{c}
0\\
- \cos \alpha\\
\sin \alpha 
\end{array}
\right)\quad \ 
\end{eqnarray}
\begin{figure}[t]
\includegraphics[width=0.49\textwidth]{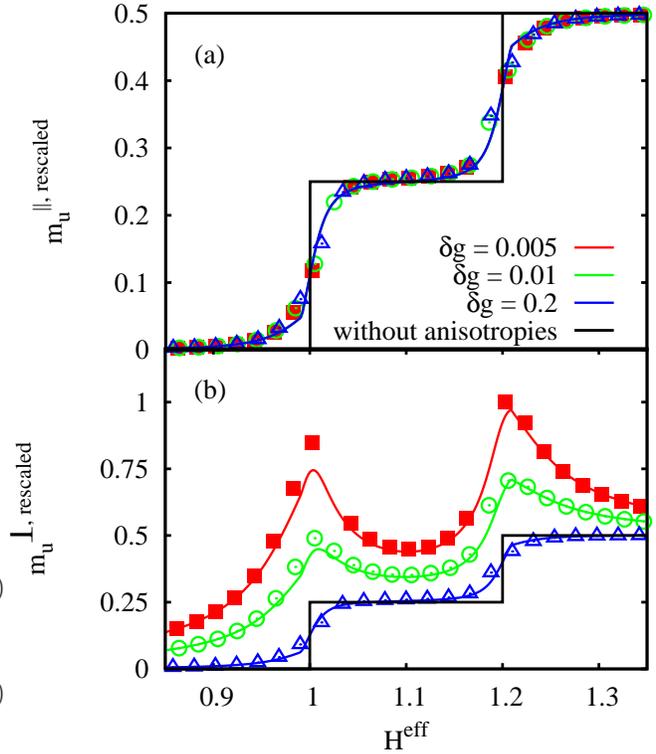}
\caption{(Colour on-line) Rescaled uniform magnetization as obtained by
  eqs. (\ref{eq:mupar_rescaled}) and (\ref{eq:muperp_rescaled})
  parallel (a) and perpendicular (b) to $\vec{H}$  for $D =
  0.03$, $\alpha = 1^{\circ}$, and $0.005 \leq \delta g \leq 0.2$. 
  The solid lines are the results of the mean-field calculation, the
  points the results of DMRG calculations on 49 rungs.
  The black line shows the exact magnetization parallel to $\vec{H}$
  when $D = \delta g = 0$.
  Note that in (a) the mean-field results lie on top of each other.} 
\label{Fig:04}
\end{figure}
Both quantities are proportional to $\left|\left\langle \vec{S}
\right\rangle \right|$ and hence $m_u^{\parallel}$ and $m_u^{\perp}$
have the same dependence on the magnitude of the field $H^{\rm eff}$, 
so that $m_u^{\parallel}$ can be obtained by measuring the torque.  
For a better comparison of the results for different parameters, we
rescale the magnetizations by their angular- and g-dependence, i.e.,
if $\delta g \neq 0$ we analyze
\begin{eqnarray}
m_u^{\parallel,{\rm rescaled}} &=& \frac{\left|
  \vec{m}_u^{\parallel}\right|}
  {\sqrt{g^2 + \cos^2\alpha \, \delta g (2g + \delta g)}}\label{eq:mupar_rescaled}\\
%  {\sqrt{\sin^2 \alpha g^2 +  \cos^2 \alpha (g+\delta g)^2}} 
m_u^{\perp,{\rm rescaled}} &=& \left| \vec{m}_u^{\perp}\right| \frac{\sqrt{g^2 + \cos^2\alpha \, \delta g (2g + \delta g)}}{\sin \alpha \cos \alpha \, \delta g (2 g + \delta g)}
\label{eq:muperp_rescaled}
\end{eqnarray}
as a function of 
\begin{equation}
H^{\rm eff} = \mu_B H
% \sqrt{g^2 \sin^2 \alpha + (g+\delta g)^2\cos^2 \alpha }.
{\sqrt{g^2 + \cos^2\alpha \, \delta g (2g + \delta g)}}.
\end{equation}
Before we go to the general case with finite $\delta g$ and finite
$D$, we first discuss the illustrative case of a system with DM
interactions only. 
To do this, it is useful to recall the results obtained for an
isolated dimer with $D \neq 0, \, \delta g = 0$\cite{PRB_Shin}.
For perturbatively weak fields one finds 
\begin{equation}
\vec{m}_u \sim \left( \vec{D} \times \vec{H} \right) \times \vec{D},
\end{equation}
leading, in general, to a magnetization perpendicular to $\vec{H}$.
Note, however, that in this case it is not possible to determine
$m_u^{\parallel}$ by measuring $\tau$, since $m_u^{\parallel}(H) \neq
m_u^{\perp}(H)$ due to the lack of SU(2) symmetry. 
This is demonstrated in fig. \ref{Fig:02} where we present DMRG
results for the case $D = 0.03$ and $\delta g = 0$. 
As can be seen, the field dependences of $m_u^{\parallel}$ and
$m_u^{\perp}$ differ fundamentally.
In $m_u^{\parallel}$ a plateau is obtained which is smoothed out due
to the DM interaction, while in $m_u^{\perp}$ pronounced peaks appear
at the critical fields confining the plateau. 
This can be understood by considering the angular dependence of the
torque $\tau(\alpha)$. 
As discussed in ref. \cite{PRB_Shin} for the case of a single
dimer, it changes drastically when leaving the limit of perturbatively
small $H$ and going to the critical field $H_c$; one finds 
\begin{equation}
\tau(\alpha) \sim
\left \{ 
\begin{array}{ll}
\sin \alpha  \cos \alpha  & {\rm for } \quad H \ll H_c\\
\sin \alpha  & {\rm for } \quad H = H_c.
\end{array}
\right.
\label{Eq:angulardependence_dimer}
\end{equation}
This leads to a clear peak in $m_u^{\perp}$ at $H_c$.
%As can be seen, $m_u^{\parallel}$ shows, as expected, clear signatures 
%of a plateau which is rounded due to the DM interaction. 
%The larger $\alpha$ (i.e., the smaller the angle between $\vec{H}$ and 
%$\vec{D}$), the clearer the plateau.
%The case $\alpha = 0^{\circ}$ is the one described in
%Ref. \onlinecite{PRL_Karlo}.
%For this case, Fig. \ref{Fig:02}(b) shows that $m_u^{\perp} = 0$.
%When increasing $\alpha$, however, the torque becomes finite, and we
%get pronounced peaks at the critical fields which grow with
%$\alpha$, as expected from (\ref{Eq:angulardependence_dimer}). 
Note that in fig. \ref{Fig:02}(b) we present $m_u^{\perp}$ rescaled by
$\sin \alpha \cos \alpha $, taking into account the angular
dependence for $H \ll H_c$. 
As can be seen, the same dependence on $\alpha$ is found for all values of $H$ sufficiently away from the critical points. 
%As can be seen, for all values of $H$ far enough from $H_c$ the lines
%collapse on each other indicating that this is the correct
%angular dependence for the ladder system when $H \neq H_c$.
In fig. \ref{Fig:03} we compare the angular dependence of
$m_u^{\perp}$ at the critical point $\mu_B \mathbf{g} H_{c,1} = 1$ and
inside the plateau for $\mu_B \mathbf{g} H = 1.1$.
In the latter case, the finite size DMRG and the mean-field results
are in excellent agreement with each other and with the dependence
$\tau(\alpha) \sim \sin \alpha \cos \alpha$, and we see that finite
size effects play a minor role.
At the critical field, however, we see that also for the ladder
systems a completely different angular dependence is obtained.
For small $\alpha$, it follows $\sin \alpha$, while for larger angles
$m_u^{\perp}$ deviates to smaller values.
Note that at the critical point the results of the DMRG and of the
mean-field calculations show qualitatively similar behavior, but the
mean-field value is smaller than the DMRG result.
Comparison of DMRG results for 25 and 49 rungs shows that finite size
effects are minimal also in this case, so that we believe that these
results should be representative for the thermodynamic limit.   
Despite the deviation from the $\sin \alpha$, for large
$\alpha$ the resulting $m_u^{\perp}$ at the critical field and inside
the plateau differ by an order of magnitude, providing an explanation
for the peaks visible in fig. \ref{Fig:02}(b).
% are due to the different angular dependence of $\tau$ in the
% vicinity and away from $H_c$.  
%This suggests that, at least for certain frustrated materials with a
%considerable DM interaction, the critical fields $H_c$ limiting
%plateaux phases can be determined with a high precision by measuring
%the angular dependence of the torque around the borders of the
%plateaux.
%The effect of the interplay of $\mathbf{g}$-tensor and DM anisotropies
%on such plateaux phases has not been investigated before. 
%Usually the assumption is made that the contributions to the torque 
%caused by D and by $\delta g$ are additive.
%One might expect that in the presence of both anisotropies torque
%measurements provide $m_u^{\parallel}$ up to minor changes due to the
%small DM contribution.
%Note, however, that 
%already for a single dimer at the critical field this is not true, as
%can be seen in first order degenerate perturbation theory.
%in general around critical fields a non-trivial field dependence is
%obtained if both anisotropies are of comparable magnitude. 
%We thus believe that our investigations on the relatively simple
%ladder system gives insights into the interpretation of torque
%measurements in general.

In the following we test how a finite $\delta g$ changes the picture
obtained for finite $D$ only. 
In fig. \ref{Fig:04} we show our results for ladder systems when
$\alpha = 1^\circ$ and $0.005 \leq \delta g \leq 0.2$, i.e. we go to
a realistic value of $\delta g = 10\%$ while keeping  $D = 0.03$. 
If $\delta g < D$, the peaks in $m_u^{\perp}$ at the critical
fields are well visible. 
%For $\delta g = 10\%$, 
For $\delta g > D$, however, the peaks vanish.
In this case, the magnetization curves resemble the 
SU(2) symmetric case up to a
smoothening of the plateau, and we 
obtain $m_u^{\parallel {\rm , rescaled}}(H^{\rm eff}) \approx m_u^{\perp {\rm ,
    rescaled}}(H^{\rm eff}) $. 

\begin{figure}[t]
\includegraphics[width=0.49\textwidth]{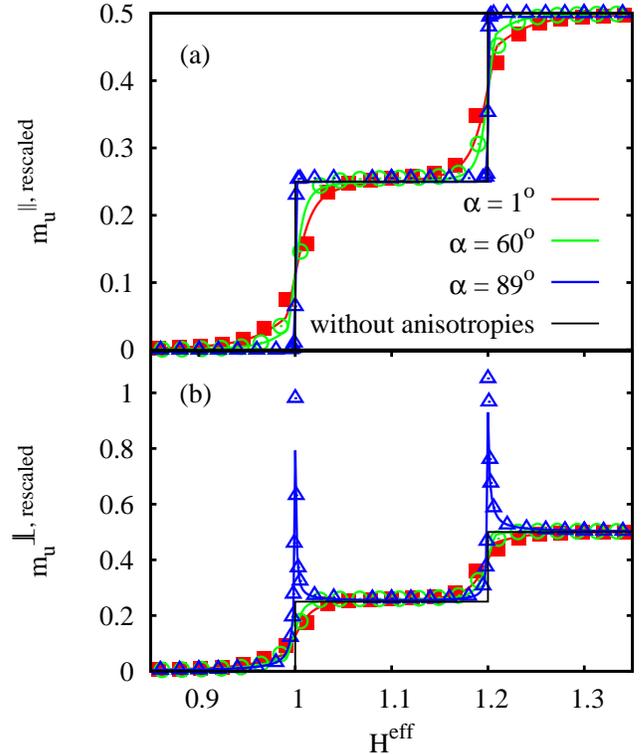}
\caption{(Colour on-line) Same as in fig. \ref{Fig:04} when varying the angle
  $1^\circ \leq \alpha \leq 89^\circ$ for $D = 0.03$ in the presence
  of large g-tensor anisotropies $\delta g = 0.2$.}  
\label{Fig:05}
\end{figure}

The remarkable angular dependence of the
torque obtained for systems with $\delta g = 0$ 
has also very important consequences in the presence of g-tensor anisotropy. 
If we now keep realistic values of both anisotropies $\delta g = 10\%$
and $D = 0.03$ and increase $\alpha$, for angles $\alpha > 80^\circ$
pronounced peaks in $m_u^{\perp}$ at the critical fields become
visible again, as can be seen in fig. \ref{Fig:05}.  
%This suggests that for frustrated dimer materials with DM and
%$g-$tensor anisotropies the critical fields can be determined very 
%accurately by measuring the magnetization with the field close to
%being parallel to $\vec{D}$.
We find that $m_u^{\perp{\rm , rescaled}}(H^{\rm eff}) \neq
m_u^{\parallel{\rm , rescaled}}(H^{\rm eff})$, as in the case with DM
anisotropy only.
This shows that care has to be taken when interpreting results of
torque measurements, and that, in general, it is not possible to
safely conclude on the field dependence of $m_u^{\parallel}$ by
considering the field dependence of the torque.

In conclusion, by investigating the interplay of DM and g-tensor
anisotropies on the magnetic response of a strongly frustrated ladder, 
we have clarified in which cases the torque  can be used as a good
approximation to the magnetization when DM is present.
For this to be the case, two conditions must be fulfilled:
i) The ratio of the DM interaction to the exchange coupling
$D/J$ should be smaller than (or at most comparable to) the g-tensor 
anisotropy;  ii) The angle between the
magnetic field and the DM vector should not be too small. While the 
first condition is often fulfilled (the DM interaction is typically a few percent
of the exchange while the g-tensor anisotropy is about 10\% for Cu$^{2+}$),
the second condition may or may not be satisfied depending on the experimental
conditions. Interestingly enough, the fact that the field dependence of the torque 
is quite different from that of the magnetization when the field is almost parallel to the
DM vector could be an advantage. Indeed, in this
geometry we predict the torque to have well pronounced peaks at
the critical fields that delimitate the plateau. This effect could thus be used
to locate with a high precision the critical fields, a difficult task if only magnetization
data are available since DM interactions can lead to a significant rounding at
the boundaries of the plateaux. In view of the controversies regarding the
plateau structure of SrCu$_2$(BO$_3$)$_2$, it would thus be particularly
interesting to perform torque measurements with the field perpendicular
to the layers and parallel (or almost parallel) to the intra-dimer DM vector
of one type of dimers, and it is our hope that the present paper will
encourage such an investigation.

\acknowledgments
We acknowledge useful discussions with C. Berthier, J. Dorier, M. Horvati\'c, S. Miyahara, K. Penc, I. Rousochatzakis,
K. P. Schmidt, I. Sheikin, M. Takigawa, and T. A. T\'oth.
This work was supported by the Swiss National Fund and by MaNEP.

\end{document}